# Retirement Transition in the Digital Ecology: Reflecting on Identity Reconstruction and Technology Appropriation

Mao Mao[a]*, Alan F. Blackwell[b], David A.Good[a]

[a]*Department of Psychology, University of Cambridge, Cambridge, United Kingdom;*
[b]*Computer Laboratory, University of Cambridge, Cambridge, United Kingdom*

*Corresponding author: Email: mm992@cam.ac.uk. Tel: +44 07934000877

# Retirement Transition in the Digital Ecology: Reflecting on Identity Reconstruction and Technology Appropriation

This paper describes a qualitative study of retirees' social and personal practices via digital music technologies in the context of community music. We conducted a diary study, and interviewed retired community musicians who are experiencing transition to retirement. Amongst challenges due to ageing and retirement, retirees participating in community music often experience discontinuity of identity caused by the lack of social and personal support after retirement, and also report lack of interest in using new technologies. Life transition theory was used to understand retirees' perception and strategies of identity navigation, informing the design of community-oriented online music services. We deepened our understanding of retirement transitions with technologies by showing how retirees participating in community music make sense of new rules and norms after retirement. A key finding is that retirees reconstruct identities by connecting with music communities, through which they can develop an understanding of unfamiliar patterns of the retired life, and gain support musically and socially. Technologies act as 'boundary objects' for communication between digital and physical artefacts, personal and social relationships. We highlight the importance of managing artefactual and interpersonal boundaries when designing online services for individual and communities in transitions.

Subject classification codes: include these here if the journal requires them

**Introduction**

Retirement from regular paid employment is an interesting point of transition in later-middle age. Retirees encounter fundamental changes in daily routines and social relationships, which will subsequently affect social and personal wellbeing. In recent years, those who are aged 65 and older, as with other age groups, are moving towards digitally connected lives (Smith, Anderson, & Page, 2017). New retirees from the 1950s

Baby Boom generation have had substantial experience of using the Internet as part of their working lives(Carr, 2010; Durrant et al., 2017). Consequently, retirees' orientation to Internet-enabled technologies is likely to change in this generation (Durrant et al., 2017). Despite this, in the field of ageing and HCI, only a few research studies have focused on *transitions into retirement*, in which retirement transitions are regarded as one aspect of the general ageing process (Creech, Hallam, et al., 2014; Salovaara, Lehmuskallio, Hedman, Valkonen, & Näsänen, 2010), or one of the many cases of life disruptions (Markus & Wurf, 1987). This suggests an opportunity to understand retirement transition in the digital ecology, whereby early retirees are faced with both new opportunities and loss of established benefits, in both social and technical aspects of their lives.

A review of existing work on later adulthood in HCI reveals that much of it is focused on an outcome-oriented perspective — oriented toward a few universal, ideal outcomes of ageing (e.g., ageing in place, maintaining continuity or staying active). A growing body of research has demonstrated the benefits and barriers of active artistic engagement for retirees (Creech, Varvarigou, Hallam, McQueen, & Gaunt, 2014; Hallam, Creech, Varvarigou, & McQueen, 2012a), as well as strategies (Hallam et al., 2012a; Hallam, Creech, Varvarigou, & McQueen, 2012b) to encourage people to get involved in communal activities. This is of course important. However, there are many situations where these universal values are problematic. In some cases, just emphasising these universal values may result in older adults experiencing ageism (Lazar, Diaz, Brewer, Kim, & Piper, 2017) or becoming isolated from activities due to poor physical conditions (M. M. Baltes & Carstensen, 1996). Researchers in CSCW and HCI have only recently begun to recognise the importance of taking a transitional perspective in studying later adulthood (Durrant et al., 2017). Drawing upon the life-span perspective, gerontologists Baltes and Carstensen (M. M. Baltes & Carstensen, 1996) argued that heterogeneity in ageing could be addressed by looking at how individuals deal with opportunities and losses. Durrant et al. noted that retirement transition around technology use has the potential impact on the construction of "sense of self", identity, and subjective wellbeing (Durrant et al., 2017). For communal activities at retirement transitions, this ongoing process of construction is just as important as the concrete outcomes of provision for ageing. Internet-based technologies should support enjoyment and sociality throughout the activity. To this end, research that paints a richer picture of the process of retirement transitions and the role of Internet-based technologies is necessary.

This richer picture of retirement transitions has important implications for research in HCI. First, the understanding of identity navigation and the re-structuring of the 'sense of self' advances research into the complexity of human developmental experiences extended through the life-course. Second, research into communal artistic activities around the use of technology among retirees speaks to several developing fields in HCI: participating in leisure activities (Lazar & Nguyen, 2017), coping with life disruptions (Massimi & Baecker, 2010; Massimi, Bender, Witteman, & Ahmed, 2014; Massimi, Dimond, & Le Dantec, 2012), and group music making.

This study contributes to the life transition research in detailing the ways that communal artistic activities take place with ecological, individual, and social factors intertwined. More specifically, this study takes a nuanced look at communal artistic practices employed by retirees around the re-construction of identity while in transitions. We draw from Gennep's work on *rite de passage* and the model of selective optimisation with compensation (SOC) as the lens through which we could describe the developing practices. While starting from a unique case of community musicians, the target participant group is middle-aged or older people who regularly attend communal artistic activities for enjoyment and learning. Our results suggest that Internet-based technologies for music play a role in connecting physical and digital boundaries, and that the appropriation of music technologies in turn assists the navigation of identities and people's perception of retirement transitions.

**Literature Review**

*Selfhood at Retirement Transitions*

Retirement transition is the process whereby people move from fully-engaged employment to retirement (Semaan, Britton, & Dosono, 2016). The study of life transition is a developing field within the ageing and psychology literature, looking at "the various challenges that people confront over the course of their lives when they encounter disruptions or change" (Hulme & Lord Wei of Shoreditch, 2014). The conceptualisation of 'retirement transition' has changed in recognising greater fluidity around this transition. With increase of human life expectancy, age is becoming less and less a determinant of retirement. Combining the processes of entering old age and leaving the workforce, retirement transition theories have been adopting a life-span view, or a "process-oriented approach" (M. M. Baltes & Carstensen, 1996). Recent research into retirement transition notes that retirement transition is an "ongoing" process that "lasts a long time" (Semaan et al., 2016), always accompanied with changes in work and life structures and mental health issues (Hulme & Lord Wei of Shoreditch, 2014). To this end, researchers believe that the core research question needs to be broadened from simply focusing on the outcomes to include "What are the processes that allow for a successful transition to retirement" (M. M. Baltes & Carstensen, 1996). To address this question of selfhood in the context of the digital ecology, we draw on two branches of theory: process models of successful ageing and stage-based transition theory.

*Process Models of Successful Ageing*

The model of selective optimisation with compensation (SOC model) (M. M. Baltes & Carstensen, 1996; P. B. Baltes & Baltes, 1990) is a process model that attempts to "represent the nature of development and ageing with the focus on successful adaptation" (M. M. Baltes & Carstensen, 1996). With a wealth of empirical evidence, successful ageing literature agrees on the notion of life-span development (P. B. Baltes & Baltes, 1990; P. B. Baltes, Reese, & Lipsitt, 1980): people take actions to minimise losses and maximise gains. The SOC model specifies three processes: selection, compensation, and optimisation. Selection refers to the increasing life constraints in anticipation of the developmental process. Compensation refers to behaviours or the acquisition of new skills to maintain the same goals in terms of resource losses or behaviour deficiency. Optimisation refers to "enrichment and augmentation" of resources in pursuit of adaptive outcomes (M. M. Baltes & Carstensen, 1996). Empirical evidence of these three processes can be classified into concerns with identity crisis and with socioemotional selectivity theory.

Identity crisis draws on Goffman's theatre metaphor (Goffman, 1959) in which identity is constructed based on rules and norms that people are engaged in. First, self-efficacy theory suggested that agency beliefs guide how the anticipatory scenarios are constructed and coped with (M. M. Baltes & Carstensen, 1996; Bandura & Jourden, 1991). The strength of self-efficacy beliefs guides how selection and compensation are determined and performed, including in the use of technologies to construct such scenarios. Second, the literature regarding "multiple selves" claims that various self-schemata (e.g., 'actual, 'feared, 'hoped for' selves) aids and guides the selection of goals (Markus & Nurius, 1986; Markus & Wurf, 1987). The selection strategy of maintaining multiple selves echoes the literature into role theory (Biddle, 2013). During life transitions, people reconstruct their self- and social- identities in pursuit of continuity in life structure and resilience, in both technologically mediated and conventional social contexts. Loss of continuity is a cause of identity crisis and psychological concerns (Osborne, 2012). Role theory argues that one way to combat identity crisis is to maintain multiple roles and identities, especially when previously established identities do not work (Biddle, 2013). Third, the social cognitive mechanism of social comparison aids the transition processes of the SOC model. In the face of losses and deficiency in retirement transitions, downward comparison may help people to maintain a positive self-evaluation (M. M. Baltes & Carstensen, 1996; Taylor, 1983). Similarly, recent HCI literature on situated learning suggests that the social context of learning serves as a significant factor in technology acceptance and collaborative use of technologies (Suchman, 1987; Xambó et al., 2013). Taken together, examining identity crisis and how people reconstruct their identities at retirement transitions is a key factor of retirement transition in the digital ecology.

There is a complementary body of ageing literature proposed to account for the three processes in the SOC model. Socioemotional selectivity theory argues that emotional satisfaction "becomes increasingly salient with age" and that retirees engage in activities that maximise their affective goals (L. L. Carstensen, 1995; Laura L. Carstensen, Isaacowitz, & Charles, 1999). The motivation of maintaining close relationships guides the selection between peripheral social relationships and closed others. In this regard, the social selection of long-term relationships helps to compensate for behavioural deficiency (M. M. Baltes & Carstensen, 1996). This optimisation of ageing and self-development is indirectly supported by empirical evidence that older couples are happier and display more efforts to combat emotional conflict (M. M. Baltes & Carstensen, 1996). Taken together, successful ageing literature advocates a process-oriented approach to answer questions about retirement transitions. By directing attention to the strategies people use to cope with life transitions or disruptions, the SOC model offers a positive perspective in which losses may lead to other gains and value in later life (M. M. Baltes & Carstensen, 1996).

*Stage-Based Transition Theory*

The second branch of theory for conceptualising retirement transitions is stage-based transition theory. In Arnold Van Gennep's theory of *rites de passage* (Van Gennep, 2011), there are three phases in a transition process: separation, liminality, and incorporation. The separation phase is marked by direct life disruption (e.g., retirement), in which people experience an initial detachment from their prior working life – in the digital ecology, this would include loss of access to devices and network infrastructure in office environments. The liminality phase plays an important part in the passage from employment to retirement. In this phase, the old identities and norms during employment may conflict with the norms of retirement. People depend on external resources, potentially including digitally accessed resources and technologies, to learn and make sense of the new rules and norms. The incorporation phase occurs when people form new habits to the new situations, after the period of combating the old and new. In this phase, new identity is formed and integrated to the new life structures. For example, becoming a formal member of activity communities is an important form of social integration (Kim & Moen, 2001b). Other stage-based models include seven stages in Erikson's developmental model that are specifically for retirement transition (Osborne, 2009), and Hershenson's five-status model (Hershenson, 2016). Hershenson describes the constructions of retirement as "any combination or sequences" of the following statuses: retrenchment, exploration, try-out, involvement, reconsideration, and exiting (Hershenson, 2016). All models are based on the assumption that experiences during prior stages would shed light on the development of personalities and identity in later stages (Osborne, 2009).

Stage-based models suggest that the characteristics of transition in the digital ecology would be defined temporally, albeit with some consideration towards the heterogeneous nature of older age that have been noted in ageing and HCI literature (Vines, Pritchard, Wright, Olivier, & Brittain, 2015). However, these models lack details regarding individual level in characterising each phase or stage. We therefore consider both process-oriented successful ageing theories and stage-based models in order to understand questions of selfhood at retirement transitions in the digital ecology.

*Personal and Social Development in Retirement Transitions*

Digital technologies potentially contribute to, or play a role in, human development during retirement transitions in many ways: the sense of self (personal), the sense of context (social and community), and resource affordance (economic and technological). The psychology literature highlights demographic and psychosocial factors that play an important role in determining how people experience retirement transitions in the digital ecology. Self-efficacy and self-esteem are important factors facilitating successful transitions to retirement (Kim & Moen, 2001b; Mutran, Reitzes, & Fernandez, 1997). For example, Charness & Boot (Charness & Boot, 2009; Hoare, 2011) found that self-efficacy can be instrumental in fostering adaptive learning of new skills among older population and as a result, increasing their subjective wellbeing. Expectations and perceptions of control has been well documented in literatures describing the relationship between individual differences and subjective wellbeing (van Solinge & Henkens, 2008).

Retirement transition, as a developmental process, also "takes place in the context of ongoing social relations" (Kim & Moen, 2001a). Technologies that support disruption of these relations have been explored in HCI and CSCW for many years. Massimi (Massimi, 2013) explored how technology could help with remembrance, social support, and bereavement by redesigning the website of an online support group. Additionally, the communication needs of retiring population are dynamic, in accordance with the diversification and contraction of social network (Heckhausen, 2001). It is likely that individuals will narrow down their social network to closer relationship (S. E. Lindley, Harper, & Sellen, 2009) and expand new social contacts regarding possible tasks in the near future (e.g., serious leisure, civic engagement) (L. L. Carstensen, 1993).

Community involvement, for example noted as a factor influencing retirement adjustment of professional women (Price, 2003), is a key question in our focus on community music and life transition. Creech and colleagues identified musical possible selves among older people and important factors contributing to better subjective well-being (a sense of purpose, a significant degree of autonomy, and a strong sense of social affirmation) (Creech, Hallam, et al., 2014). Overall, involvement in music communities is found to be associated with improved subjective well-being (Hallam, 2010; Hallam et al., 2012a, 2012b). Built on the conceptualisation of sense of community (D. W. McMillan & Chavis, 1986), community involvement influenced identity formation and development in life transitions in four ways: membership and rituals, in-group influences, integration of needs, and shared emotional connections. In the case of music, (Creech, Varvarigou, et al., 2014), *Creech et al.* identified key practices of community music (e.g., scaffolding, modelling, giving and receiving feedback) that contributes to the identity navigation and interpersonal relationship development. While these findings are important, more research is needed to unpack how social and personal development of retirement population is grounded in communities.

Prior to the emergence of ICTs, the social and personal perspectives of retirement transition were confined to physical environment and offline communications. Today, however, it comprises social interactions and self-representations via various technology-mediated communications tools and offline networks. To this end, we are interested in examining the online and offline practices of people undergoing a transition in the context of community music.

*Technologies provide opportunities for communities*

Research on the role of technologies during life transitions has explored cases including residential moves (S. Lindley & Wallace, 2015), breakups (Moncur, Gibson, & Herron, 2016), domestic violence (Massimi et al., 2012), remembrance and bereavement (Massimi, 2013), homelessness, job loss, post-disaster technology adoption (Shklovski, Burke, Kiesler, & Kraut, 2010), and the transition from military to civilian life (Semaan et al., 2016). These studies highlight how people reconstruct social-digital infrastructures in support of transition.

An important part of later life transitions within the digital ecology is the adoption of new technologies. In examining this, researchers have highlighted that technology adoption needs to fit into the specific social and cultural context, and the needs of people during transitions shifts across time. Shklovski et al. (Shklovski et al., 2010), in a longitude study of musicians in the aftermath of Hurricane Katrina in New Orleans, found that musicians adapted computer-mediated communication and information seeking tools for personal and community use. Lindley and Wallace (S. Lindley & Wallace, 2015) highlighted the importance of continuity in combating changes in residential environments for older people, and recommended that technology could focus on supporting meaningful spending of time currently and prospectively. Both studies stated that online technologies allow the older generation to regain control over their personal and social life by connecting with others and seeking information online.

HCI research concerning major identity change over time has addressed the effect of social computing technologies on other transition processes. Haimson et al. (Haimson, Bowser, Melcer, & Churchill, 2015) studied the relationship between Facebook account management strategies during gender transitions and its psychological effects, finding that online social network could mitigate the stress by utilising a less granular infrastructure and more user control. In other work looking at the process of finding a new normal after war, Semaan et al. (Semaan et al., 2016) discussed the role of ICTs across the three stages of transitions from veteran to civilians. Brewer and Piper investigated the identity construction among older adult bloggers, suggesting that life logging systems such as blog supports the development of identity, helps with self-expression, and enables a sense of belonging and social connectedness (Brewer & Piper, 2016). However, despite the potential of these identified practices involving life transitions, a more nuanced look towards different events and cases is necessary to better support the developmental process of transition as occurring in retirement transition.

With regard to the role of communication technologies during life transitions, regular contact is the heart of interpersonal relationship (Burke & Kraut, 2014) and community involvement. O'Flynn identified four types of Facebook communications among college choir members: sharing of information, commentary after performances and other events, sharing of music files and general expressions pertaining to everyday events and/or personal feelings (o'Flynn, 2015). Massimi and Neustaedter (Massimi & Neustaedter, 2014) conducted research on people's use of video chat to support the sharing of major life events. They identified use patterns of sharing major life events between close others across long distances; and the ways in which atmosphere and ambience of sharing influence experience of these practices. Given that a large body of work in HCI and CSCW has explored remote communication among families and friends, this existing work highlights the value of understanding how ICTs serve communication needs within communities.

Research has addressed the interaction with physical and digital artefacts across lifespan to uncover the technology ecology of life transitions. We draw on theoretical constructs of Artefactual Intelligence by *David de Léon*, which is an ecology of human achievements concerning information flow among people, digital and physical artefacts, spatial conditions, and time (De Léon, 2003). In HCI research concerning life transitions and processions, *Moncur* raised several issues related to digital assets among people at the end of their lives, during which memory, identity, and reputation entwined with each other (Moncur, 2015). In an interview study (S. Lindley & Wallace, 2015) looking at moving to care homes in later life, Lindley and Wallace noted that music streaming services could mitigate the effect of relinquished possessions by either providing digital file archiving or being regarded as a compensation strategy for loss. Aside from research for the older population, HCI and CSCW researchers have also studied the role of technology during relationship breakdown (Moncur et al., 2016), sharing photos of baby on social media among mothers (Kumar & Schoenebeck, 2015), short-term and long-term curation of social media content (Zhao & Lindley, 2014).

Finally, research prototypes have been studied to make sense of the ecology of artefacts management. In the work of Gulotta et al. (Gulotta, Sciuto, Kelliher, & Forlizzi, 2015) and Merritt et al. (Merritt, Jones, Ackerman, & Lasecki, 2017), tools were designed to curate digital collections with metadata. Among identified person-generated and system-generated factors, time is regarded as a contextual factor in shaping the ecology. A relatively small number of projects have addressed design of technology to manage music among older people: for example, McMillan and colleagues designed a tangible music player prototype called Pick Up and Play (D. McMillan, Brown, Sellen, Lindley, & Martens, 2015), with which users can copy music, curate playlists and explore new music by simply manipulating small cubes. Our own research focuses on understanding issues related to the context and requirements of such technologies in future, through better understanding of existing technology appropriation among this population today.

**Method**

*Diary-aided Interview*

We conducted the study from January to June of 2015. An advertisement was distributed among community choir members living in our location in Eastern England. At the end of the interview, the participants were thanked and given a CD of choral music for taking part.

Prior to diary-aided interviews, we built rapport and trust with participants via in-person meetings, and in many cases joining the group rehearsals to sing with them for one or two weeks. Before the interview, participants were first asked to keep a diary over the course of one week. The diary structure consisted of six days of conventional diary taking, followed by a drawing task. In each of the first six days, the participants were asked to write down their music-related activities, technologies that they have used, and the context of using the technologies. This was done three times each day – in the morning, afternoon, and evening. The task on the seventh day is a drawing task, in which the participants were asked to draw a "social diagram" regarding all their social connections of music, and in what way they were connected. Here participants may write down the names of their connections in the choir, family, and acquaintances. For example, Figure 1 shows a reproduced social diagram drawn by a participant attending community music. This participant wrote "Me" in the centre of the diagram, and connected herself with communities that she was in via music instruments and sheet music. Some participants wrote or drew music-related technologies such as radio, laptops and MP3s used for maintaining the social network of music.

Figure 1. A reproduced social diagram from one participant

Participants were then invited to our research lab for a 60-min semi-structured interview on their experience of doing community music, and how they use and perceive digital music technologies. Participants were first asked to explain their social diagrams and the reasons behind them. We asked who the people or organisations were, how the mentioned technologies were used, and how the diagram developed over time. We also encouraged participants to talk about how they manipulate and experience offline music (such as radios, vinyl, and CDs) and to compare these experiences with their use of digital music technologies. All interviews were conducted in person and audio-recorded.

*Participants*

Eleven people participated in the diary-aided interview, 3 men and 8 women with ages ranging from 50 to 71, from Eastern England. They had been recruited via connections of the research team, distribution of posters and flyers, and the mailing lists of local community choirs. Our participants have diverse backgrounds of employment status: in employment, semi-retired (taking on one or several part-time jobs while being retired from a full-time job), and fully retired (from one month to ten years).

Our participants have diverse educational backgrounds and jobs (including music-relevant as well as unrelated jobs). Two of them had experience of leading a choir, teaching and conducting. Recruiting both community music leaders/conductors and members gave us insights into diverse strata of this community, and how digital music technologies are used and perceived in different contexts. Importantly, six of them have used some types of digital music services (e.g., BBC iPlayer, iTunes, SoundCloud, Spotify). The other five were asked to experiment with such services during the interview in order to capture their initial reactions. A description of the participants' demographics is provided in Table 1.

Table 1. Demographics of participants

*Analysis*

The interviews were all audio recorded and transcribed. We performed qualitative data analysis using a Grounded Theory Method (GTM) (Strauss & Corbin, 1990). At first, two researchers did the open coding of all transcripts independently, during which memos were written. Then the two researchers iteratively discussed the memos and codes, and performed axial coding and formed categories. Finally, theories were generated from data. The themes reported are the major ones regarding main personal and social practices surrounding community music. Many of the practices were analysed in terms of life transition as discussed earlier.

**Findings**

We begin by drawing nuanced definitions of community music from our data, describing how people develop practices in terms of three different phases of transition to retirement: separation, liminal, and incorporation phase. All respondents reported the tension between the prospective self and the actual self at the transition. Through our analysis, we found that those who participated in community music developed ICT-mediated daily practices while participating in communal artistic activities.

*Characterising the life at transitions with community music*

Our data indicates a distinction between two levels of standard of music aspiration in community music: *formal* groups and *informal groups*. In formal groups, musicians regularly perform choral pieces mainly for a paying audience at formal venues. In informal groups, people with the common interests in music participated in a range of communal music activities mainly for the enjoyment of music. Note that informal group members normally have a relatively lower level of musical aspiration compared to formal music groups, even if some informal music groups perform at formal venues for a paying audience. The difference lies in the levels of complexity and texture of choral works.

Our interviewees mentioned various issues related to retirement transition with community music. The most commonly mentioned challenges are about knowing who they are and finding new purpose in life, which is guided by and intertwined with the type of music groups that retirees belong to. The distinction between formal and informal groups has implications for the membership and rituals of community music practices. Drawing upon research about the sense of community, membership of a group is formed with established boundaries, a sense of belonging, and identification(D. W. McMillan & Chavis, 1986). Community music groups create boundaries with the rituals of music – learning music, rehearsing and performing musical pieces, and interacting with audiences. As a result, individuals form their identifications and share emotional connections that are specific for the group that they belong to. For instance, six participants mentioned their experience of attending workshops – a type of short-term informal music group. The enjoyment of attending informal music groups is often linked to the experiences of being in a new place, having a nice performance venue, meeting new people, and interacting with audiences and other musicians. One of the interviewees described her experience of singing in an informal choir, in which novel performance venues and the shared emotional connections were mentioned:

> "It was such fun! It was lovely to sing on the steps of [name of a museum] and they [the cycling athletes were] all going past […] The sort of social network, its choir leaders, workshop leaders, friends [attracted me most], […] And the leader at the workshop in Letchworth, she is the most well organised person." (P02)

This contrasts with formal group, an example of which comes from a participant who had shifted from a higher standard formal group to an informal one. To him, the ritual of formal music groups was less flexible, as it prevented him from singing during off-term periods when no rehearsals were scheduled. The higher level of standard of music at the formal choir resulted in pressure on him.

> "I count my main choir as C [choir], and I feel less stressed, M [choir] was a bit stressful, when I was so new to singing, but there is [not] a lot of pressure on, as I went to a really good performance [in the C choir] – I loved it. But it is[has] less pressure". (P10)

In this case, although retirees are able to find the music groups that are appropriate to their identity and needs, the identities and needs at life transitions change over time. Just as P10 found choir M a good one for him to become a better musician in the first place, a more important need of gaining social interaction and relaxing time drives him to switch to choir C with less demanding work. In contrast with informal music groups, the sense of belonging to a formal music group is likely to be associated with more advanced musical skills, higher level of standard of music, and more nuanced musical performance, in addition to the general enjoyment, commitment, and social support.

Moreover, retirement transition is associated with the early process of ageing, and the need for appropriation of Internet and streaming technologies makes the situation more complex than purely musical accomplishment. All our participants mentioned lack of confidence or lack of interest in new technologies. For recent retirees, not being able to update the knowledge of technology is a main reason for this:

> "I am hopeless on computers. I learnt from it when I was working. Since when I stopped working, I want to move them away […] Not have [I] been brought up with a computer technology, smartphone, etc. So I tend to stick to what is familiar, like radio." (P03)

Even for those who have substantial knowledge of technology, activities in which digital technologies (i.e., transferring music from CDs to the MP3 devices) are involved are always regarded as something "mentally demanding" and are seldom regarded as relaxing or of most interest:

> It would be quite nice to have my iPod with the classic music that I have […] I really would like to have new music [on the iPod], and that would be just a process of transferring [music] from CD to the iPod. But I found it [requires] quite [a bit of] mental effort to work out. And I need sort of a nice, clear day, like today, and while I [have] a feeling [of] clear-headed, and I've got nothing that was worrying me. Then I could sit down and face that." (P06)

The characteristics of retirement transition that we identify here are broader and more specific than presented in prior work, which has not focused on the implications that relatively informal and formal communities might have on identity and the appropriation of digital technologies. We unpack the experience of retirement transition based on Gennep's transitional theory, what implications it has for the sense of community, and how different phases of retirement transitions are situated in informal or formal community music practices.

### *Separation from the working life*

Why do people join community music groups at the transition to retirement? We found that, for those who had never done community music before, participating in communal music activities was motivated by either personal interest in music or the desire to seek new social connections. For those who have been doing music all the time, joining or re-joining community music acted as a strategy to maintain their prior musical selves.

Our participants talked about how their social relationships were manifested with music collections and digital technologies when they separated from the life of employment. Our analysis shows that "downsizing", "revisiting", and "the redefinition of self" are three primary themes in this phase.

#### *Downsizing music possessions*

Our results indicate that participants tend to get rid of their music possessions that are "inappropriate or even overwhelming" (S. Lindley & Wallace, 2015) during times of retirement transition, and only leave those manageable or compatible with their personal tastes. As retirees experience the transition from work to retirement, their perception of time and space change. The ways in which the downsizing of music possessions is performed are not consistent across community musicians, and did not necessarily entail the trend of technology progress, either physical or digital.

When given an option, the strategy of downsizing and applying technical skills attained before retirement were intertwined. The decision whether to retain or discard certain types of music collections was influenced by their perceived control in relation to these music formats. For example, one participant (P10) preferred buying CDs from ChoraLine, which is a website providing resources of learning choral music, instead of downloading music from the website. This is because CDs are more manageable than digital files for him. He also mentioned that he "has got more than enough (music) to keep (him) going for the rest of his life". For him, music possessions enable the formation of the future-facing self. That is to say, music collections are not only being viewed as "triggers for memory", but also as "possessions in sustaining particular practices now and in the future" (S. Lindley & Wallace, 2015).

In accordance with Lindley and Wallace (S. Lindley & Wallace, 2015), downsizing is seen as a way to support existing relationships and interests of our participants. Around the age of formal retirement, people are more likely to focus on maintain close relationships rather than increasing their social networks (S. Lindley, Harper, & Sellen, 2008). In our data, gift receiving and giving were typical examples of downsizing practices at retirement transitions. Three participants reported receiving and giving CDs as gifts upon retirement, and were thinking of giving some of their CDs away after retirement. For example, P06, a semi-retired, 71-year-old female, keeps the physical music collections related to choral singing. Choral singing is a major part of her current life; she would like to buy more when needed. Whilst at the same time, she has been thinking of giving away a large collection of CDs and vinyl left by her father, as "it is not my taste of music". This points to the gap existing in current HCI research that advocates design of systems to support reminiscence in older adulthood: the way that retirees deal with music possessions is compatible with their future-facing selves, and focusing on reminiscence as a priority for technology use by older adults would potentially impede the possibility of personal and social development mediated by technology.

Of course, our participants do collect musical artefacts for the purpose of reminiscence. One typical example is saving programmes, flyers, and newspapers related to music groups and performances. This practice is generally being replaced by digital technologies. As participant 10 put it, "[keeping and reading the past music programmes] might be interesting but I can find it out on a computer." To this end, the affordances of digital technologies and Internet-based music services would offer further opportunities of reminiscence in relation to community music.

Downsizing is not without challenge. Our participants experienced tension when their needs did not align with a technical solution, e.g., managing physical music collections and the increasing scale of digital music metadata, and the intertwined use of digital and physical music devices. For example, participant 04 usually passed her books onto her friends as a gift. However, what made her recently annoyed was the difficulty in giving her Kindle books to her friends as a gift:

> "You buy something comes on to Kindle, but you can't give it to somebody away afterwards, these stuffs on your Kindle, not allowed. [Even if you delete the books from the device] they still sit up there on cloud." (P04)

In this example, the cloud platform of Amazon Kindle meets the need for downsizing of physical book collections, and in the meanwhile mediates the access to shared possessions of book, and this was also the case for shared music possessions. In the first place, for recent retirees in our study, the knowledge of cloud technology and shared repositories are relatively lacking. This further adds to the complexity of sharing digital files among retirees. Secondly, the meaning of shared repositories matters

(Gruning & Lindley, 2016). Giving a Kindle book away is different from sharing a Kindle book: the former points to the transferring of ownership of the digital book, whilst the latter suggests the collective ownership of the digital book. Obviously, the current affordance of digital sharing technologies is not explicit in meeting such needs. In this sense, it is necessary to design meaningful structures that retirees can understand.

*Revisiting recording of music*

Another typical community music practice among retirees is to revisit what individuals already own. For retirees doing community music, this means revisiting music collections, in order to organise, replace, discard older collections, to upgrade existing devices/technologies, and to learn from different interpretations of music.

First, participants in both formal and informal music groups are motivated to revisit their existing music collections when they are facing the situation of selecting collections and format of music to be retained. Possible reasons for doing so include re-organising collections at the workplace and at home right after retirement, and responding to the limited space when moving home. For example, one participant who just retired one month ago was reaching the point of moving books and document from his prior office to his home, and he had to negotiate with his wife about making room for his music collections. He noted that *"…we are reaching the point that we couldn't fit any more in the space, we have to start selecting."*

In addition, participants in both formal and informal music groups would revisit their music collections while upgrading their music storage technologies, for instance, transferring music from CDs to minidisks. This process is long and tedious, but in the meanwhile, streaming services are found useful for this practice:

> "Likely the process of moving to minidisks forced me to listen to everything that I've got. This is because of comparing. Sometimes I consciously want to retain a number of the recordings, better go every variation, probably six recordings. And I do have a favourite, but I play the others from time to time. Because there are different interpretations, but as I said earlier, Spotify enables me to do that." (P05)

In our data, specifically, formal music group singers often revisited music in support of learning performance repertoire. The analysis of the diary study showed that all participants reported learning music in the performance repertoire at least once in the week of completing the diary book. There are two types of music recordings in support of learning music: recordings created by instructors, and music pieces recorded by musicians themselves. On the one hand, music facilitators and choir leaders shared recordings of performance among choir members via email or the choir website. Participants learned music by listening to these and other recordings, which they accessed in two primary ways: (1) by searching for recordings saved on their personal computers (via file names searches of the melody name) and (2) by searching for a CD title in an alphabetical index. As participant 02 noted:

> "She [the choir leader] names it [the music files]. They [the music recordings] are already named with the song […] I did [revisit the downloaded music recordings]. I actually save my music file section on my laptop, which I am proud of. They are all there, you just need to [save them] because when [we] did a concert in December in [name of a cathedral] and we were singing songs that we learnt from another one [choir] in summer […] Yes, I keep them. I actually save them." (P02)

On the other hand, our participants recorded the rehearsed songs by themselves and revisited the music in person or in groups. In addition to learning, revisiting these recorded songs brings enjoyment and shared memories:

> "We were learning songs all the time, and some people recorded them right to the end when we went through what we've learnt. On the last night, we went through what songs we have learnt in the two weeks. It was really lovely, really nice." (P03)

Revisiting digital and physical music is a typical practice at retirement transition, through which our participants separate from their formal identities as an employee. Digital music files and devices function like a bridge between the online and offline possessions. Our participants negotiated new spaces and upgraded their technological devices. Particularly for those with higher levels of musical aspiration, revisiting music recordings provided a foundation for enriching a musical-self that had previously been underdeveloped.

*The redefinition of self*

The third, and no less important theme that emerged through our analysis is that many of our participants reported their act of re-organising life, work, and social network for the purpose of "finding new ways of spending time" (S. Lindley & Wallace, 2015), while separating from their prior identities. In the process of separating, some participants left work and moved to a different place. After retirement, some participants found that their prior sense of being at the centre of work, groups, and communities no longer applied:

> I have always been in a way of [sort of] central figure, within the church community […] But when you retired, you are no longer in that central position, nor do I necessarily want to be, but you do [need to] make friends." (P08)

The direct consequence of retirement motivated our participants to seek new activities to fill their time and adapt to the change. In our data, participation in community music was one such activity. On a deeper level, joining community activities enabled participants to redefine their identity. Many participants emphasised the importance of keeping busy after retirement and using new activities to establish new future-facing identities. In our data, future-facing identities were framed in two ways.

Firstly, future-facing identities are framed via managing possessions of music. In the case of participant 06, her willingness of having more choral music collections rather than those from her father demonstrated her current and future-facing identity in connecting to community music and spending time and money on music consumption. Here we see that the future-facing strategy of downsizing is bound up with the willingness to "build new social networks and finding new ways of spending time" (S. Lindley & Wallace, 2015).

Moreover, identities are framed with music genres, singers, and communities. When faced with the decision making of which type of community to join and which type of community music to focus on, respondents' 'music selves' become critical. Personal preferences for music genres, musicians and the atmosphere of music communities are among the top criteria in the selection of music groups to join: "I love folk music, and I think I can sing along too. Thus, I chose this choir [name of the choir]." (P03)

To separate themselves from their prior life and habits, our participants seek out ways to redefine their self-identities by "downsizing" and "revisiting" music possessions. These acts help retirees adapt to the forthcoming new routines, as well as get them prepared for potential adjustment. In this sense, what existing HCI and gerontology research is worrying about, the retrenchment of space, finance, and social network at retirement transitions, need not be a cause of concern for a successful transition.

### *Liminality in artefacts and identities*
When people dispose of an old way of life, they move into an objectification phase – a very confusing time in which old habits no longer work and new patterns of behaviour have not yet been created. The transitioning retirees in our study who were in this stage gathered information, thought about possible activities and styles of community music, and selected and tried out these activities and styles. All of our participants reported that they struggled to use new ICTs when exploring and trying out community music activities.

On the one hand, our participants were reluctant to ask for help from their family because they did not think their family members (i.e., their children) had sufficient time or patience. Further, the participants did not feel confident that they would be able to apply their existing technological skills to post-retirement life: "I just can't remember it [my password]. I am getting on [with using smart phones]; I am getting on. It is not a thing that I take to easily at all" (P09).

On the other hand, we found that the participants navigated to Internet-based media sharing services in support of group music making and received support from their community music friends in using new technologies. Here, we report how retirees developed music learning and sharing practices to make sense of their new routines post-retirement. We also describe their motivation to use new technologies in order to become a better musician and a successful retiree.

*Identity navigation through learning and sharing*

Our participants succeeded, in many cases, in creating their own identity support patterns. To navigate various confusing selves at times of transition, participants stepped into several online-offline sharing activities and collaboration practices, such as using recording devices to record music sessions and using email and cloud services to share sheet music, recordings, and photos. Once our participants were successful at connecting with these tools and resources, they began to expand their identity support network. Community music-based websites, such as ChoraLine (see Figure 2), enabled community musicians to share and receive information, and get support from choir leaders. As described by P03:

> "We joined with other choirs, and they have a big mass choir. And [the choir leader] put those songs on Dropbox, so we can access them, and learn the songs. Because we don't actually learn them in classes, we do it by ourselves at home. And then we get together for a couple of weeks, or sessions, before our performance, [and] go through them." (P03)

Figure 2. A screenshot of www.choraline.com

Music shared through cloud services and email attachments was mainly used in two ways. First, it was used to support learning and performance. The retired musicians were self-motivated to interact with the music files and sheet music to better engage with their groups. The community music instructors and conductors distributed this content, providing resources and tools to all of the community musicians. Such music sharing drove participants to try new ICTs. For instance, P11 made full use of new technologies to support her work as a conductor and choir instructor.

> With my performing choir, somebody records the session every week, and I put it onto Dropbox […] I also, for the choir, usually we record the separate parts [of music] and put them [the media files] on Dropbox, so people can listen to that part and all the parts together. (P11)

Moreover, shared music recordings are used to establish collective identities and attract audiences and attendees. Group websites, social network pages, streaming services, and email newsletters are used for distribution. Some participants cited such websites or pages when discussing the style of their music groups. This illustrates that the websites and services that supported music sharing were extensions of collective identity.

The formal and informal music groups used the shared music recordings and sheet music in different ways. In formal groups, recordings were weakly associated with sharing – musicians seldom recorded rehearsals by themselves or shared recordings with their groups. In contrast, informal groups frequently made recordings and shared these in order to assist members with learning the music outside of rehearsals. Sheet music was shared with the expectation that it would be printed out. Just as we found in the previous study (Mao, Blackwell, Lukate, & Good, 2016), print sheet music was the dominant medium of musical interaction, in both formal and informal groups. Two participants mentioned that they "printed the sheet music out" after receiving it via email or a cloud storage system, in preparation for the community music sessions. This finding necessitates the consideration of the different materiality of the shared artefacts in community music and the way in which people perceive and use them.

*Being motivated to use ICTs: The effect of self-efficacy*
In our data, participants were motivated to use technology for three reasons: (1) they had recently bought/obtained a device; (2) they had been influenced to do so from people in their community; or (3) they had help available at hand.

On the one hand, our participants became ICT users at various stages of the retirement transition when the technologies were perceived useful and helpful in assisting their musical selves. For example, P05 became satisfied with streaming music services after having fixed the problem of connecting his old (but still good) sound system to his streaming devices:

"I've got Spotify there and then the Sonos, [but] I can't use it now. What it does is that it controls the speakers in my house through which I can stream Spotify or radio to the speaker. It [the Sonos] has a unit, which allows me to stream [music from Spotify], and [recently] it is saying "I can't find the system". I think it [the Sonos] is a wonderful thing. In my living room, I have another thing, which is called the Gramofon: it is a little box plugged into my amplifier and that allows me to stream through my old 30-year-old sound system. I felt that the Sonos had a nice sound, but not really [as] the quality with the old system. Now someone told me about the little thing, which costs about 40 pounds. You just plug it into the back and then, you are using your own system with the quality of the old system, fantastic!" (P05)

This above statement points to a recurring theme that the materiality of technology and life transitions intertwined in the context of community music (Faraj & Azad, 2012). Some music consumption experience turned digital (e.g., use of an online music library) while the rest was highly dependent on the physical devices (e.g., a traditional sound system). Retirees such as P05 perceived music streaming technologies as useful, specifically when the new technologies were compatible with traditional forms of technologies. Thus, to address retirees' "lack of interest" towards novel technological systems, we should make the affordance of new technologies explicit, and consider both and contextual dimensions in their development. For example, as noted by P05, music streaming services provide the explicit benefits of compatibility with traditional sound systems.

Moreover, participants were motivated to learn new technologies when they perceived the technologies as beneficial for their engagement in community activities. When participants gained facility with such technologies, they felt a sense of mastery and continued motivation to engage with those technologies. For example, the motivation of one participant's use of YouTube was influenced by the social network of music.

> "I was forced to learn how to use YouTube. Because the choir people keep saying, "Oh there is a wonderful recording on YouTube, and listen to that", whereas before I had never watched [videos on] YouTube." (P04)

However, use of new ICTs as an identity navigation strategy also came with challenges. With limited knowledge of cloud services – or limited support from family and friends – some respondents found it difficult to continue with using ICTs, and this difficulty might have prevented them from participating in social sharing in communities:

> "Somebody has sent me something, in fact it was from the choir, with pictures from Dropbox, and I set it up, and then I got it a real mess and I deleted it [Dropbox]. And I decided that I don't want to be in that [using Dropbox], and then somebody else send[t] some photos from Dropbox, and I thought, I would quite like to see them but [using Dropbox is not easy for me]. And then I am not on Facebook, either, I refused to go on that." (P04)

The effect of self-efficacy during the retirement transition has been documented by prior research (Donaldson, Earl, & Muratore, 2010; Hershenson, 2016). In our study, participants' motivation to use ICTs to engage with music and navigate their identity was inevitably framed by their confidence in learning and using technologies. Per our analysis, the impact of self-efficacy was different in formal and informal music groups. In formal music groups with higher musical aspirations, participants focused on growing their personal musical expertise and paid less attention to using of ICTs for social interactions. Music sharing practices mainly occurred through more traditional forms of interacting, such as experiencing physical music together and sending files via email, which demanded less self-efficacy with technology. In contrast, informal group members tended to frequently share and receive recorded music pieces to sing along with, and have conversations about music, both online and offline. Informal group members interact more frequently with music sharing technologies. The more they interact, the more they will develop a shared music history and are more likely to become closer. Such practices sometimes demanded greater confidence in using new technologies. In other words, technology-specific self-efficacy is a more important motivating factor for social interaction rather than personal skills development among informal group members.

In sum, Internet-based and digital music technologies played an important role in influencing how our community musicians navigate identity at the retirement transition.

### *Incorporation: Life re-structuring and aspirations*

At the incorporation phase of a transition, an individual accepts that the transition is final. In our study, participants reported the experience of finally re-structuring their lives after exploring and trying out of different post-retirement choices. Only at this time did the new situation finally seem comfortable. Similar to the findings of previous studies (Semaan et al., 2016), our study found that participants ultimately got used to living with their newly formed self-identity as a retiree. Six of the participants began to engage in music sharing practices with music streaming services and cloud technologies during this phase.

#### *Finding meaningful ways to engage in community music*
The community musicians in our study reported learning how to act in the retired world and finding acceptance as a senior citizen. Retirees who participated in community music generally became used to the uncertainties of their new life rebuilding their "music-specific" social network. Participants perceived this phase to be meaningful and developed a sense of belonging to the group they were currently involved in.

One form of meaningful involvement related to finding one or several appropriate groups to join. For instance, after trying out many music groups of different professional levels, P08 finally found the one that he felt most comfortable with.

> "I find that [name of the choir] suits me well. I can't imagine that I have a [more ambitious] choral future in the sense that it's going to get very much better than what I've got, I try to go to as many common things [choirs] as I can" (P08)

P08 was aware of the change when he had experienced when shifting from a more advanced chorus programme to his current one:

> "My choral ambitions will be declining after all that wonderful experience. I am in my 60s now and in my 70s, it would probably be a downhill from here […] I count my main choir as [name of the choir], and I feel less stressed". (P08)

P08 embraced the incorporation phase when he found a sense of belonging in his choir. There, his musical-self mirrored to the level of musical aspiration of the group. In addition to the standard of musical aspirations of particular communities, voice repertoires and performance culture also played an important role in participants' decisions to participate in one or more groups. Retirees based decisions not only on the extent to which a certain group matched their self-identity, but also the way in which they presented themselves to group members. Our participants developed a "participation strategy", which they compared to those of other group members. Although P09 found performing in a concert hall as part of a formal choir attractive, she also enjoyed taking part in a less professional and smaller choir:

> "The acoustic [at W choir] when we are singing are not great…[but] I can hold my line and can come with the right note. It helps the other people who don't sing at all, apart from the [name of the choir] which is very ad hoc." (P09)

P09 "develop[ed] her self-worth" (O'Bryan, 2015) in the choir by presenting herself in a role that "giving" supported to her fellow singers. As Lave and Wenger (Lave & Wenger, 1991) put it: "the value of participation to the community and the learner lies in becoming part of the community". To this end, forming a meaningful self-identity and social membership in community music helped participants find a sense of meaningful involvement.

Transferring and sharing knowledge from a different domain was another strategy for meaningful involvement, both personally and socially. Using new technologies, our participants adapted their music-specific identities to the broader retirement context. On a personal level, P08 used a master's-level course and a three-month sabbatical during his 50s to learn effective use of ICTs. These technical skills proved additionally beneficial in helping him find information, search for activities, learn music using online resources, stay connected with friends/family and community members, and develop a sophisticated perspective of social media. He "now uses a computer quite a lot", and had "access to a lot of things that he has never ever thought [he] would have before".

On a social level, informal technical help within the community was mentioned by three participants. For example, P11 was happy to provide technological equipment to one group member who was a great help to her in editing music files with a professional music editing tool:

> "I have somebody in my choir. He has a free space – he doesn't pay to come. As an exchange of that he is the Sibelius[1] man. So if I've got something, I send it to him.

---

[1] Sibelius is a commercial scorewriter software developed by Sibelius Software. It is used by music practitioners. The main functions of Sibelius include editing/printing scores, playing music back, and producing legible scores for editing, printing and publishing.

He can get it mixed, he could get it not for money, but as a skill. And he likes to do that." (P11)

In a French translating group at the University of the Third Age (U3A), P04 used a simple tapping move that she had learnt from friends in another group to successfully navigate a Kindle dictionary during translations. From the above cases, it is evident that participants' technical skills helped them improve the effectiveness and enjoyment of their participation in community groups, either by increasing their self-efficacy or improving their social capital within and across these groups.

*Reconsideration and Exiting: the multifaceted retirement transition*
Our participants reported a variety of issues related to their presentation and perception of themselves in the community, as well as when should they reconsider their membership in this community.

**Changing identities over time.** The community musician's sense of self may change in response to their vocal expressiveness over time. As an "expressive human activity" (O'Bryan, 2015) rooted in all cultures, singing has the function of communicating human emotions, needs, and meanings. In our study, the choral singers felt that their voices defined them, and the loss of voice that can accompany the typical ageing process was a source of distress. Three of our participants had experienced this issue. As a result, they had to reconsider their participation in the voice group:

> "And there is a point about 20 years ago, I had a cancer treatment, giving therapy actually takes away three notes. [Before the treatment] I was the higher soprano […] and when I went back to [name of the choir], […] I stayed for a year [as an alto]. It [singing as an alto] was an interesting thing to do because [my voice was] in the middle of the massive sound, and I happened to be a bit more on the board about sight-reading – I actually quite enjoyed that. […] then my voice sort of improved and I am second soprano now." (P10)

Restructuring the self in the incorporation stage could be addressed musically or non-musically, personally or socially, offline or online. For P10, the temporary leave from the soprano part entailed feelings of loss, but also celebration: she experienced a growth in her personal musical skill levels, and this surprisingly improved her self-efficacy with music. In another case, (the oldest) participants were found to have left music groups as singers, as they had decreased their self-efficacy with music, but to have re-joined as a special audience. Feeling of anxiety and guilt over not being as competent as others were cited as participants' main reason for exiting the music groups. However, these participants celebrated this exit as the start of another status with another hobby (e.g., involvement in taking care of guide dogs) or a family-focused lifestyle. As indicated by Hershenson ( 2016), "a person can be in multiple statuses" at the same time. Our findings are compatible with this, as the ending of one activity involves an embrace of new activities and continued navigation of the self.

**Tensions between public and personal presentation.** While the idea that persons navigate self-identities is not new (Semaan et al., 2016), the idea that digital technologies might facilitate persons' presentation/perception of self-identities (e.g., via music streaming services) seems novel, particularly in the context of retiring community musicians. In our study, retirees strategically perceived others' online identities and built their own identity. P07 mentioned the tension she experienced when discovering that a video of her had been posted on YouTube by one of her choir friends. Her first reaction was surprise: "I was not told that I was to be put up on YouTube, so it was a bit naughty." Later, she found the humour in it and told a few friends to look it up. However, she had concerns over the accessibility of YouTube among friends of her age: "I don't have people who had ever looked at it, and [my friends] used [YouTube] rarely." Later, this situation triggered new concerns about her self-presentation online and her desire to control her own privacy:

> "I don't know if YouTube makes things down after certain time or something. I would not have chosen […] if I was going to pick a video or even audio of mine, singing to a public domain, this is not the situation I would have chosen. I've got much better recordings of that" (P07)

The participant explained that her tension came from the mismatching between her existing online self-presentation as an academic staff member and the funny video with poor recording quality: "if you Googled me, maybe [you] want to know about my professional activity or something you've got with me in evening dress singing an old-fashioned pop song." For P07, firmer boundaries between her public and personal self-presentation would have given her a stronger sense of control over her online disclosure. This sense of control was prominent for the community musicians, especially when the transition from offline to online self-presentation occurred without their knowledge. This raises important questions about the way in which music streaming services might better support the region for personal and closed social circles, as content shared via streaming platforms passes first pass through the public space unless specific privacy management actions are conducted.

Another form of transition that related to participants' public-/self-presentation was the development of a more sophisticated mental model of others' online identities of others. Participants reported using various heuristics to make sense of other musicians' skill levels on YouTube. Two indicators were used to determine the quality of online music search results. One was the age of the performer: the younger the performers, the less expertise of music they were deemed to hold. Participants tended to link videos with younger performers to:

> "Poor quality recording, poor quality performance and lots of American students doing final year recitals or music colleges like that, doing video and putting them up on YouTube […] Quite a lot of them are not good." (P07)

P05 provided further justification:

> "All sort of [music] has been done by very bright young kids who made enormous progress technically but they haven't got any music in them at all. [These recordings] lack musical expression." (P05)

The second indicator of perceived music quality related to performance venue. Our participants only deemed performances in well-known venues and by trustworthy orchestras of high quality:

> "Sometimes you can get a clue perhaps as the place that they are singing from: if this is somebody who is singing from Vienna or Paris, it is much more likely to be of higher standard, because they've got so much competition. Whereas If it is somewhere in the middle of nowhere, it is more likely to be a local person who is really proud of what they are doing, but [the quality is] perhaps not very well" (P05)

Overall, music streaming services provide an important platform for community and social interaction amongst the retirees participating in community music. For these participants, finding meaningful ways to get involved – by selecting the most suitable activities for their online and offline identity – drove personal growth and social capital within the communities. This further motivated those in the transitions to retirement to continue to contribute and share, both online and offline, with their communities.

**Discussion**

In this paper, we have explored the retirement transition in three phases: separation, liminality, and incorporation. Our research tackled two challenges faced by retirees participating in community music: navigating identity and lacking interest in adopting digital technologies. The findings suggest that digital music technologies play an important role in supporting the navigation of identity and that they are appropriated in the context of community music. Technologies that support downsizing, revisiting, learning, and sharing music could be potentially useful in addressing retirees' lack of interest in new technologies. In the following section, we draw on our data to reflect on the implications of this research.

### *Retirement as a process*

The overarching theme of this study and related work is that the retirement transition is a process that incorporates the selection, compensation, and optimisation of resources (i.e., information technology), and a time at which retirees reconstruct their identity to reach an emotionally salient goal. This study enhances the life transition literature by providing a rich and detailed case of community music. Drawing upon Van Gennep's (Van Gennep, 2011) transition model, we identified representative social and personal practices in the context of community music and described how these practices shift across the retirement transition process. We also explored the way in which individuals make sense of the new rules and norms during this process.

Revisiting and downsizing music possessions are typical practices in the separation phase of the retirement transition. In this phase, the emotionally salient goal is to enact a new self-identity that is adaptable to changes. Although perceived constraints and resource strategies differ between individuals (relating to, e.g., lack of space at home, moving one's home/office, and the availability of new technologies), our participants performed practices such as downsizing and revisiting their music collections in order to adapt to new circumstances. In the liminality phase, learning and sharing practices are typical optimisation strategies for maintaining autonomy in daily routines and social relationships. Our participants acquired new skills (both social and musical) that assisted their adaptation in the retirement transition. In the incorporation phase, participants became used to the new norms and rules of retirement, though their reconstruction of identities did not end. Selection strategies such as trying out new activities and reconsidering the appropriateness of current identities, in addition to compensation strategies, such as exiting of current communities, led participants to the optimised outcome of staying positive in later life.

In contrast to identity reconstruction during general life transitions, identity reconstruction during the retirement transition requires people to both "deal with the prospect of change" (S. Lindley & Wallace, 2015) and cope with many other transitions that occur along with retirement. However, we might still ask: Rather than supporting a certain type of transition, how can we take a broader scope to support identity reconstruction during life?

### *Assisting offline practices with online technologies*

The core of identity reconstruction is to make sense of the "big picture" – the new rules and norms – in which people are embedded (Semaan et al., 2016). In this study, we identified two types of boundaries across the three stages of transition: artefactual and interpersonal boundaries.

In our analysis, physical and digital artefacts in the context of community music were bound up with personal and community practices. Downsizing and revisiting – two essential community music practices in the separation phase –highlighted the importance of negotiating the value and meaning of physical music collections and digital music formats. First, the affordances of music streaming services were found to mitigate the mental burden of "relinquished" music collections (S. Lindley & Wallace,

2015) and to facilitate access to more music. As in the case of P05, technologies (e.g., Spotify, the connector between the laptop and the sound system) that integrate his 30-year-old sound system perfectly met his music listening needs at home. This case highlights the opportunities for new technologies to connect with existing physical music devices (e.g., old speakers with good quality) and music collections. Second, the tension that participants associated with gifting physical and digital music artefacts speaks to the notion that digital metadata predicts user intentions. For example, artefact-generated metadata (e.g., music genre) is important for owners when deciding whether to retain or gift music collections; in contrast, user-generated metadata (e.g., time) is only useful when owners are determining whether the music can be shared. Difficulties relating to when and how one should gift music possessions reveal future opportunities for design.

In the analysis, we found that digital music technologies acted as possible 'boundary objects' (Star & Griesemer, 1989) – objects that communicate between the retirees' family social networks and their community social networks. Specifically, a collective interest in music allowed retirees to participate in community music practices and, through this participation, enact their self and collective identities to support a successful transition to retirement. For example, our participants re-appropriated music streaming services, cloud services, and social network services, and adopted tools that had been developed for community music in order to more directly engage with community music. Meanwhile, community music practices mediated by ICTs further shaped participants' perception of interpersonal boundaries and relationships. First, the practices of music sharing mainly occurred within the social circles relating to music. Specifically, shared music files, sheet music, and even concert tickets acted as the negotiating artefacts between music and non-music social networks. Second, emotional attachment to music artefacts and emotional responses to music were bound up with basic social motives (Loersch & Arbuckle, 2013) and re-shaped social relationships by increasing social bonding (Pearce, Launay, & Dunbar, 2015). This finding is in line with findings relating to the construction of sense of community: rituals, membership, emotional attachment, and the integration of needs (D. W. McMillan & Chavis, 1986). Third, community music practices occurred in an enriched learning environment, in which people who were mainly technology novices naturally appropriated new technologies to support their routines, rather than being "assisted" in this process. On the one hand, music instructors and technology savvy administrators distributed music for learning, mediated challenges that choir members were faced with when downloading and listening to shared media files. Their instructions and support facilitate the learning of technology (Burt, 2009). On the other hand, social support from peers affirmed the perceived usefulness of the technologies, which is attributable to technology acceptance.

*Rethinking how we evaluate the meaningful use of technologies*

The findings of this study prompt us to rethink our evaluation of the meaningful use of technologies among retirees during times of transition. Prior research on social media has addressed the redefinition of social media amongst the older population (Hope, Schwaba, & Piper, 2014). The literature on technology use among the senior population has mainly focused on the amount of use and technology acceptance at a basic level. From the perspective of technology appropriation, patterns of technology use among retirees at life transitions are likely to be different from those of the general population. The social aspects of technology use can be better understood by examining the strength and closeness of communication with community and family ties, the depth and frequency with which technology is spoken about, and the complexity of mental model of the knowledge/information-sharing network in the context of community activities. From the perspective of identity, the presence of both digital and physical formats of artefacts represents engagement in online or offline practices, respectively, and the formation of one's self-identity. A more precise description of technologies that are meaningful for life transitions (occurring both materially and temporally) is needed.

*Design implications*

Designing for retirement transitions is challenging, because transitions are multifaceted, and people undergoing life transitions are heterogeneous and always changing. Age is not the only dimension of the retirement transition. New technology design should also consider practices and routines, including the ways in which these change over time and through various media. Our analysis identified two key issues to consider when designing new digital technologies for individuals and communities experiencing in the retirement transition.

First, such technologies should enable the *presentation of identity at different levels*. Echoing past research on the communication and relationships in older age (Hope et al., 2014; S. E. Lindley et al., 2009), our findings suggest that direct and in-depth communication is best for relations in family and in retirement communities. Designers should consider the different roles that a retiree may take in the context of community music and how these roles might change over time. New technologies should facilitate users' meaningful representation of their selves and accommodate a degree of personalisation.

Second, new technologies should have perceived benefits and engender a sense of control. The findings of this study suggest that *reconsidering the affordance of technology* might improve its perceived value and enhance users' interests towards technology. For retirees, the perceived value of a piece of technology lies not only in the device or tool, itself, but also in the mutual relationship between artefacts and contexts. Verbal communication (e.g., email, text messages, and phone calls) compensate for "lightweight" communication on social media (Hope et al., 2014). Designers should also consider how technological systems might be connected or integrated with physical artefacts, so that the mutual relationship is supported within the context of use.

**Conclusions**

This paper has presented findings from an in-depth qualitative study of community musicians making the transition to retirement and the role of technology in this transition. Typical community music practices were identified across different phases of the retirement transition. This study replicated previous HCI and gerontology research on the multifaceted and dynamic nature of the retiree population and contributed a more specific set of qualitative findings about the retirement transition experience. The findings point to retirees' attempt to reconstruct identities and to acquire skills with new technology in order to adapt to changes. Digital music technologies were used in the context of community music as "boundary objects" that bridged artefactual and interpersonal boundaries. This study opens up opportunities for more and broader conversations regarding social, personal, and temporal aspects of the retirement transition. We hope an understanding of the multifaceted nature of the retirement transition among community musicians and their use of technologies will inform the design of community-level technologies that will support retirees' acquisition of new skills, sharing of digital and physical artefacts, and the representation of identities at multiple levels.

We thank all the participants and choir coordinators for sharing their thoughts. We thank Johanna Lukate for her help with coding the interview transcripts.

Table 1. Demographics of participants

| P# | Age | Employment | Individual experience of transition into older age |
|---|---|---|---|
| 01 | M(64) | Semi-retired | Doing part-time work, singing in a local informal choir, attending workshops sometimes |
| 02 | F(61) | Retired | Participating in one informal choir and a formal choir, attending workshops sometimes |
| 03 | F(67) | Retired | Singing in two informal choirs and one formal choir, used to be teaching assistant of music classes |
| 04 | F(71) | Retired | Singing in a formal choir, used to be the head of music in a primary school |
| 05 | M(66) | Retired | Retired one month ago; singing in four formal choirs |
| 06 | F(71) | Semi-retired | Singing in five formal choirs, singing as soloist in two choirs, violin player |
| 07 | F(62) | Semi-retired | Singing in one informal choir |
| 08 | M(71) | Retired | Previous church musician, choir singer in one formal choir, attending workshops sometimes |
| 09 | F(67) | Retired | Singing in two formal choirs |
| 10 | F(69) | Retired | Previously sang in seven formal choirs, mainly attending one currently |
| 11 | F(50) | Not retired | Community music practitioner, composer, and conductor |

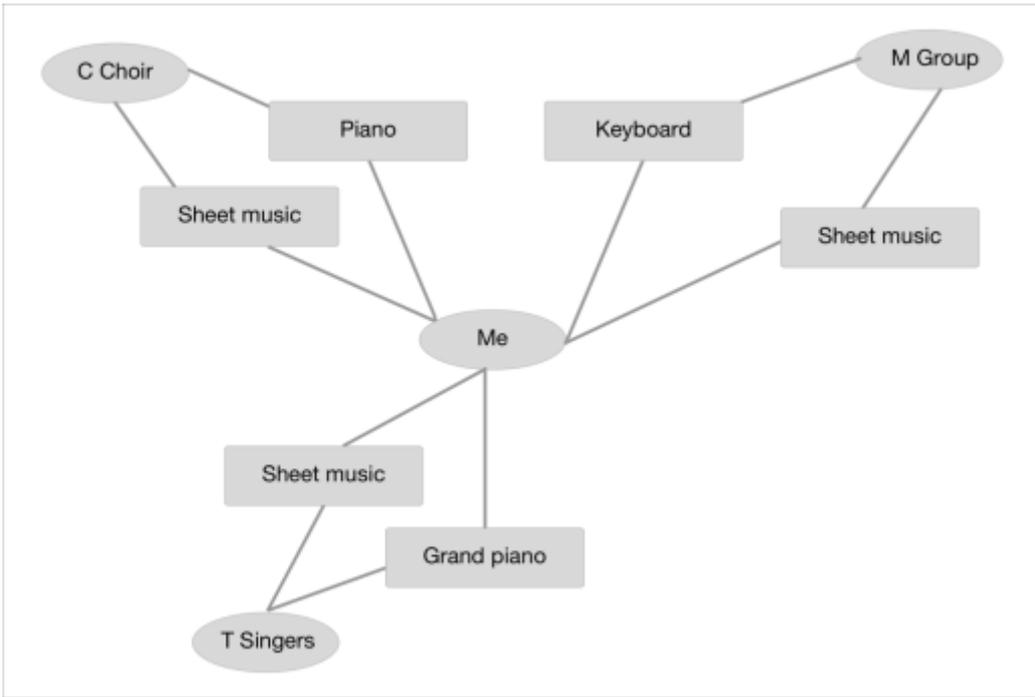

Figure 1. A reproduced social diagram from one participant

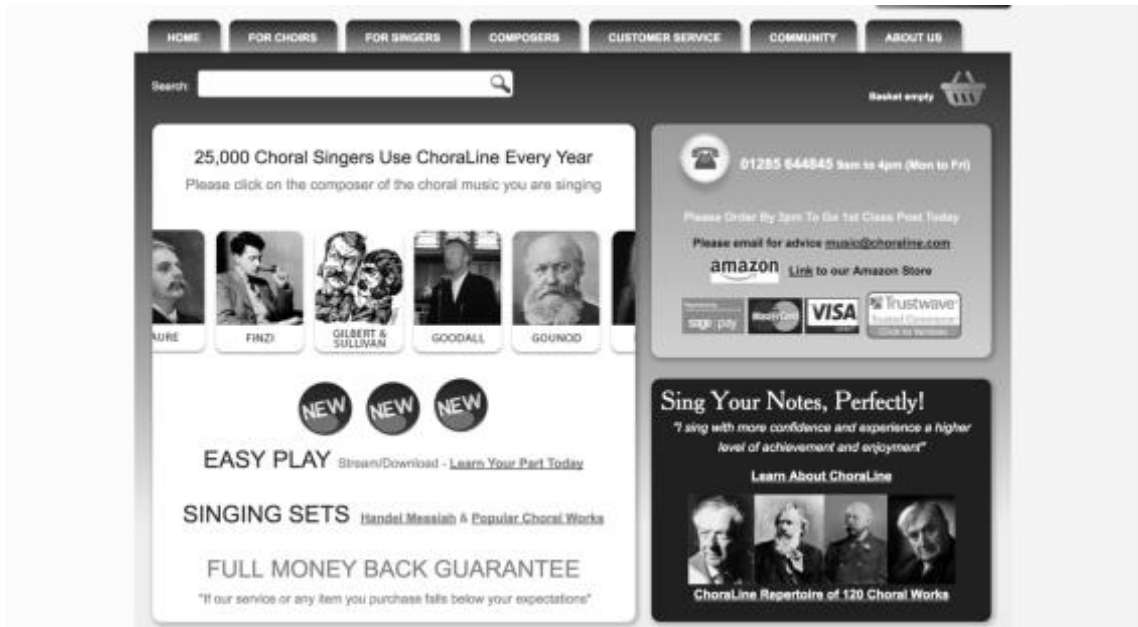

Figure 2. A screenshot of www.choraline.com